\documentclass{article}

\usepackage{amsfonts}
\usepackage{amsmath}
\usepackage{amsthm}
\usepackage{amssymb}

\hoffset= - 1.0in
\newcommand{\M}{\mathcal{M}}
\newcommand{\N}{\mathcal{N}}

\newcommand{\PS}{\mathcal{P}}
\newcommand{\map}{\rightarrow}

\newcommand{\NN}{\mathbb{N}}
\newcommand{\RF}{\mathbb{R}}
\newcommand{\dif}[2]{\dfrac{d #1}{d #2}}
\newcommand{\pdif}[2]{\dfrac{\partial #1}{\partial #2}}
\newcommand{\pdifd}[3]{\dfrac{\partial^2 #1}{\partial #2\partial #3}}

\newcommand{\bd}[1]{\left\{#1\right\}}
\newcommand{\br}[1]{\left(#1\right)}

\newcommand{\intg}[1]{\mathop{\int}_{#1}}

\newcommand{\summ}[2]{\mathop{\sum}_{#1}^{#2}}
\newcommand{\intgd}[2]{\mathop{\int}_{#1}^{#2}}
\newcommand{\la}{\lambda}
\newcommand{\ar}{\alpha}
\newcommand{\be}{\beta}
\newcommand{\g}{\gamma}
\newcommand{\om}{\omega}
\newcommand{\gatr}{\dot{\gamma}}
\newcommand{\fht}{\dot{h}}

\newcommand{\eq}[1]{\begin{equation} #1 \end{equation}}
\newcommand{\eqm}[1]{\begin{multline} #1 \end{multline}}
\newcommand{\eqlm}[2]{\begin{multline} \label{e:#1} #2 \end{multline}}
\newcommand{\eqmn}[1]{\begin{multline*} #1 \end{multline*}}
\newcommand{\eqn}[1]{\begin{equation*} #1 \end{equation*}}
\newcommand{\eql}[2]{\begin{equation} \label{e:#1} #2 \end{equation}}
\newcommand{\eqs}[1]{\begin{align} #1 \end{align}}

\newcommand{\en}[1]{\begin{enumerate}  #1 \end{enumerate}}

\newcommand{\pt}{\dot{p}}

\newcommand{\xt}{\dot{x}}
\newcommand{\yt}{\dot{y}}
\newcommand{\xtt}{\ddot{x}}

\newcommand{\ttm}{\widetilde{T\M}}
\newcommand{\ytt}{\ddot{y}}
\newcommand{\pxt}{\dot{p_x}}
\newcommand{\pyt}{\dot{p_y}}

\newcommand{\leg}{\mathfrak{L}}

\newcommand{\re}[1]{(\ref{e:#1})}
\newcommand{\tg}{\widetilde{\gamma}}

\newcommand{\s}{\sigma}
\newcommand{\ttkm}{T^{2k-1}\M}
\newcommand{\tkm}{T^{k-1}\M}
\newcommand{\omp}{\om_{\PS}}
\newcommand{\gp}{\g_{\PS}}
\newcommand{\gpt}{\dot{\g}_{\PS}}
\newcommand{\Hom}{\mathrm{Hom}}
\newcommand{\ve}{v}
\newcommand{\rit}{reparametrization invariant }
\newcommand{\x}[1]{x_{(#1)}^i}
\newcommand{\y}[1]{y_{(#1)}^i}
\newcommand{\defeq}{\displaystyle \mathop{=}^{def}}
\newcommand{\gdi}[1]{\displaystyle \mathop{\g}^{\scriptscriptstyle(#1)}\hskip -2 pt {}^i}
\newcommand{\hdi}[1]{\displaystyle \mathop{h}^{\scriptscriptstyle(#1)}\hskip -2 pt {}^i}

\DeclareMathOperator{\sign}{sign}

\allowdisplaybreaks[1]

\numberwithin{equation}{section}
\begin{document}
\title{
\begin{flushright}
 \mbox{\normalsize ITEP/TH-59/07}
\end{flushright}
\vskip 20pt
Geometric Hamiltonian Formalism for Reparametrization Invariant Theories with Higher Derivatives
}
\author{Petr Dunin-Barkowski\footnote{E-mail address: barkovs@itep.ru}\\Alexei Sleptsov\footnote{E-mail address: sleptsov@itep.ru}\\ \\ ITEP, Moscow, Russia}
\date{}
\maketitle
\begin{abstract}
Reparametrization invariant Lagrangian theories with higher derivatives are considered. We investigate the geometric structures behind these theories and construct the Hamiltonian formalism in a geometric way. The Legendre transformation which corresponds to the transition from the Lagrangian formalism to the Hamiltonian formalism is non-trivial in this case. The resulting phase bundle, i.e. the image of the Legendre transformation, is a submanifold of some cotangent bundle. We show that in our construction it is always odd-dimensional. Therefore the canonical symplectic two-form from the ambient cotangent bundle generates on the phase bundle a field of the null-directions of its restriction. It is shown that the integral lines of this field project directly to the extremals of the action on the configuration manifold. Therefore this naturally arising field is what is called the Hamilton field. We also express the corresponding Hamilton equations through the generilized Nambu bracket.
\end{abstract}

\section{Introduction}
In the Lagrangian formalism dynamical systems can be conditionally divided into the following classes:
\en{
\item non-degenerate, i.e. with $\det\br{\pdifd{L}{\dot{x}^i}{\dot{x}^j}} \neq 0$
\en{
\item $L = L(x^i,\dot{x}^i)$,

\item $L = L(x^i,\dot{x}^i,\ddot{x}^i,...)$;
}
\item degenerate, i.e. with $\det\br{\pdifd{L}{\dot{x}^i}{\dot{x}^j}} = 0$
\en{
\item $L = L(x^i,\dot{x}^i)$,

\item $L = L(x^i,\dot{x}^i,\ddot{x}^i,...)$.
}
}
The Hamiltonian formalism for these systems is well-known (see, for example, \cite{Arn,Morozov,SovGem,Tyutin}). The Lagrangian and the Hamiltonian formalisms have geometric interpretation based on symplectic geometry (see, for example, \cite{ArGi} for a treatise on symplectic geometry and \cite{Rubtsov} for some modern applications). Nowadays geometric approach in mechanics is under active study \cite{Miron,DoSto}.

Among all degenerate systems there is a specific case: reparametrization invariant systems. Such a class of systems is of interest because reparametrization invariance is a type of gauge invariance which emerges in describing relativistic particles or strings. For some interesting ideas on reparametrization invariant field theories see \cite{FGM}.

Conventional geometric methods of classical mechanics cannot be readily applied to reparametrization invariant systems. The main problem is as follows. Due to the reparametrization invariance there exists a continuous family of the initial conditions for the Cauchy problem to the Euler-Lagrange equations, every point of the family defining the same extremal curve on the configuration manifold. This fact is in some sense a degeneracy. One would like this degeneracy to vanish. For this reason we construct some space $\PS_m$ over every point $m$ of the configuration manifold $\M$. There should be, roughly speaking, a one-to-one correspondence between all the extremal curves passing through the point $m$ in $\M$ and all the points in $\PS_m$ projecting to $m$. We refer to the union $\PS = \displaystyle \mathop{\cup}_{m \in \M}\PS_m$ as the \textit{phase bundle}. On $\PS$ therefore there should exist a field, which integral curves project to extremal curves of the action on $\M$. It is called the Hamilton field. A map from the space of the initial conditions for the Cauchy problem to $\PS$ is called the Legendre transformation. The conventional formulae for the Legendre transformation (see \cite{SovGem,Tyutin}) do not satisfy our requirement: they do not eliminate the above degeneracy.

In this paper we present well-defined formulae for the Legendre transformation which satisfy the above requirement. The Hamilton equations are presented too. All these are obtained via geometric approach. The main idea which we use is that every curve has a distinguished parametrization: \textit{parametrization by the action along the curve}. The resulting formulae for the Legendre transformation are obtained in a way which in a sense resembles relativistic mechanics.

In our construction the Hamilton field arises naturally from the geometric properties of $\PS$ as the null-direction field of the symplectic two-form. The formulae for it are expressed using the generalized Nambu bracket (for information on the Nambu brackets see \cite{Takhtajan,CurtZach}).

Our paper is organized as follows. In Section \ref{ris} we discuss the reparametrization invariance. To familiarize the reader with the construction, in Section \ref{1case} we discuss the simplest and well-known case (which is also described, for example, in \cite{Rovelli}) of the systems with the first derivatives. Then, in Section \ref{gencon}, the general construction is represented. After this we outline the construction in detail for systems with the second derivatives in Section \ref{2case}. Section \ref{nambu} is somewhat stand-alone: it gives the description of the relation between the symplectic form null-vectors and the generalized Nambu bracket used in previous sections. We also provide a set of explicit examples.

\section{Reparametrization invariant systems}
\label{ris}
In this section we discuss general properties of reparametrization invariant systems. We start with the standard case of Lagrangians depending on the first derivatives only. Then we describe the general situation.

\subsection{Reparametrization invariant systems with the first derivatives}
We shall now determine what we imply by ``reparametrization invariant Lagrangian system''.

Consider a differentiable manifold $\M$ and a function $L$ defined on the tangent bundle to $\M$, i.e. $L : T\M \map \RF$. We will refer to $\M$ as the \textit{configuration manifold} and to $L$ as the \textit{Lagrange function}. Let us denote the manifold resulting from exclusion of the null section from $T\M$ as $\ttm$. Then $L$ should be a smooth function on the manifold $\ttm$ being continuous on the null section. $\M$ with $L$ together are called the \textit{Lagrangian system}.

Let $\g:[a,b] \map \M$ be a smooth curve in $\M$. Consider the integral
\eq{
S(\g) = \intgd{a}{b}L\Big(\g(t),\gatr(t)\Big)dt,
}
where $t$ is the parameter along the curve and $\gatr(t)$ is the tangent vector to the curve at the point $\g(t)$. This integral is called the \textit{action} on the curve.

A Lagrangian system is called \textit{reparametrization invariant} if the action does not depend on the parametrization of the curve, but depends only on the curve as a set in $\M$ with fixed orientation. In other words, let $\g_1:[a,b] \map \M,\; \g_2:[c,d] \map \M$ be two arbitrary parametrized curves in $\M$ such that $\g_1(a)=\g_2(c),\,\g_1(b)=\g_2(d),\,\g_1\br{[a,b]}=\g_2\br{[a,b]}$. That is, $\g_1,\,\g_2$ are two parametrizations of the same curve in $\M$ as an oriented one-dimensional submanifold. Then, the theory is called \rit if for all such curves the equality $S(\g_1)=S(\g_2)$ holds. With a slight abuse of terminology, we will refer to Lagrangians, corresponding to \rit systems, as \rit.

Note that reparametrization invariant Lagrangians resemble differential one forms in some sense. Indeed, the differential one-forms on $\M$ are just functions on $T\M$ which can be integrated over curves on $\M$ with integral independent on parametrization. The difference is that the differential one-forms are linear functions if restricted onto $T_m\M$, while the \rit Lagrangians are not. However, we shall see below that, being nonlinear, they nevertheless catch one of the properties of linear functions.

Let us find out what restrictions on $L$ does the reparametrization invariance apply. Let $\g_1:[a,b]\map \M$ be a parametrized curve. Fix a smooth monotone increasing function $f:[a,b]\map \RF$. Let its image be a segment $[c,d]\subset \RF$. One can obtain a parametrized curve $\g_2:[c,d]\map \M$ in the following way: $\g_2(\nu)=\g_1\big(f^{-1}(\nu)\big),\ \nu \in [c,d]$. That is, $\g_2$ is reparametrization of $\g_1$ with the function $f$.
Now we write expressions for the action on both of the curves:
\eqs{
&S(\g_1) = \intgd{a}{b}L\br{\g_1(\mu),\dif{}{\mu}\g_1(\mu)}d\mu,\\
&S(\g_2) = \intgd{c}{d}L\br{\g_2(\nu),\dif{}{\nu}\g_2(\nu)}d\nu = \intgd{f(a)}{f(b)}L\br{\g_2\big(f(\mu)\big),\frac{1}{f'}\dif{\g_2}{\mu}\big(f(\mu)\big)}df(\mu) =
\intgd{a}{b}L\br{\g_1(\mu),\frac{1}{f'}\dif{}{\mu}\g_1(\mu)}f'd\mu.
}
Because, by construction, the curves $\g_1,\,\g_2$ coincide as sets in $\M$, i.e. $\g_1([a,b])=\g_2([c,d])$, the reparametrization invariance condition implies that $S(\g_1)=S(\g_2)$. Therefore
\eq{
\intgd{a}{b}L\br{\g_1(\mu),\dif{}{\mu}\g_1(\mu)}d\mu = \intgd{a}{b}L\br{\g_1(\mu),\frac{1}{f'}\dif{}{\mu}\g_1(\mu)}f'd\mu.
}
Because $f$ is an arbitrary function and $\g_1$ is an arbitrary curve, we obtain the following relation for the Lagrangian:
\eq{
L(m,v) = \ar L\br{m,\dfrac{v}{\ar}},
}
for all $m\in \M,\ v\in T_m\M, \ar > 0$. Redefining $\ar$ as $\dfrac{1}{\ar}$ and writing in coordinates, one obtains
\eql{1repinvcond}{
L\br{x^i,\ar v^i} = \ar L(x^i,v^i),
}
where $x^i$ are some coordinates on $\M$ and $v^i$ are coordinates on $T_m\M$, naturally induced from $x^i$.

This means that $L$ is a degree-one homogeneous function of velocities. This property can also be written in the form of the Euler equality
\eq{
L = v^i\pdif{L}{v^i}.
}
Therefore in spite of not satisfying one of the conditions of linearity, $L(v_1+v_2)\neq L(v_1)+L(v_2)$, the Lagrangian satisfies the other: $L(\ar v) = \ar L(v)$. Hence, it turns out that only this condition is really important for the one-form to be integrated over a curve. Thus, a reparametrization invariant Lagrangian may be thought of as some sort of ``nonlinear differential one-form''\footnote{However, the Lagrangians which resemble one-forms most of all, i.e. which have the property $L(\ar v)=\ar L(v)$ not only for $\ar > 0$ but also for $\ar <0$ (for example, $L=\sqrt[3]{x^3+y^3}$), require special consideration: the corresponding Legendre transformation, which will be discussed below, does not distinguish orientations of curves. We will not discuss this subject in the present work.}.

\begin{small}
\paragraph{Example.} The most familiar example of the reparametrization invariant Lagrangian is probably given by that of the length of the curve in Euclidean space. In two dimensions in coordinates it takes the form

\eq{
L = \sqrt{\xt^2+\yt^2}.
}
\end{small}

\subsection{Reparametrization invariant systems with higher derivatives}
\label{HD}

To define the Lagrangian formalism with higher derivatives we review the notion of the $k$-th order tangent bundle to a differential manifold first.

Consider a manifold $\M$ and a fixed number $k\in \NN$. Fix a point $m\in \M$. Consider all parametrized curves on $\M$ which pass through $m$. We now define an equivalence relation on them.

We say that two curves $\g_1:[a,b]\map \M$ and $\g_2:[c,d]\map \M$, $\g_1(\mu_0)=\g_2(\nu_0)=m$, are equivalent if in some coordinate chart on $\M$ all their derivatives of all orders up to the $k$-th one coincide in the point $m$. It is clear that this relation is really an equivalence relation and that it does not depend on the chosen coordinate chart.

The set of equivalence classes of the above relation is called the \textit{$k$-th order tangent space to $\M$ in the point $m$} and is denoted by $T^k_m\M$. The spaces $T^k_m\M$ taken in all points of $\M$ together form the fiber bundle $T^k\M$.

Thus, the phrase ``the Lagrangian depends on derivatives up to $k$-th order'' really means that it is defined on $T^k\M$.

So, we call a \textit{Lagrangian systems with $k$-th derivatives} the pair $\br{\M,\ L}$, where the function $L:T^k\M\map \RF$ is smooth on $\widetilde{T^k\M}$. By $\widetilde{T^k\M}$ we imply the manifold resulting from exclusion of the null section from $T^k\M$. The null section of $T^k\M$ is defined simply as the classes of equivalence of curves with all the derivatives vanishing in the corresponding points. Note that, however, the space $T^k_m\M$ is \textit{not} a vector space for $k>1$ in contrast with the case of $k=1$.

The action for systems with higher derivatives is defined as:
\eq{
S(\g) = \intgd{a}{b}L\big(C_{\g(t)}(\g)\big)dt,
}
where $C_{\g(t)}(\g)\in T^k_{\g(t)}\M$ is the class of equivalence of the curve $\g$ in the point $\g(t)$.

Reparametrization invariant Lagrangian systems with $k$-th order derivatives are defined in the same way as in the case of the first derivatives. The system $\br{\M,\ L}$ is called \textit{reparametrization invariant} if the action on the curve does not depend on its parametrization or, in other words, for every two parametrized curves on $\M$ coinciding as oriented one-dimensional submanifolds of $\M$ the action is the same.

In the same way as was derived relation \re{1repinvcond} for \rit Lagrangians with the first derivatives, one can obtain similar relations for Lagrangians with higher derivatives. For example, in the case of the second derivatives we end up with the relation
\eql{2repinvcond}{
L\br{x^i,\ar v^i,\ar^2w^i+\be v^i} = \ar L\br{x^i,v^i,w^i},
}
where $(x^i,v^i,w^i)$ are coordinates on $T^2\M$ induced naturally from some coordinates $x^i$ on $\M$ and $\ar > 0,\, \be$ are arbitrary constants. The coordinates $v^i$ correspond to the first derivatives, and $w^i$ to the second ones.
This condition can be rewritten as a set of two equations
\eqs{
L = v^i\pdif{L}{v^i} + 2 w^i\pdif{L}{w^i},\\
v^i\pdif{L}{w^i}=0.
}
These conditions are called the \textit{Zermelo conditions} (see, for example, \cite{MasCor}).

Note that, as was mentioned, the space $T^2_m\M$ is not a vector space. Thus, the expression $L\br{x^i,\ar v^i,\ar^2w^i+\be v^i}$ may seem ill-defined because, generally speaking, in this case the point $\br{x^i,\ar v^i,\ar^2w^i+\be v^i}$ might depend on the choice of coordinate chart on $\M$. Indeed, consider two points $\xi_1,\, \xi_2\in T^2_m\M$. Introduce on $T^2_m\M$ some coordinates $(v^i,w^i)$ induced from the coordinates $x^i$ on $\M$. Let $\xi_1,\, \xi_2$ be expressed in these coordinates as $(v^i_1,w^i_1),\, (v^i_2,w^i_2)$. Then, because $T^2_m\M$ is not a linear space, the expression $(\ar v^i_1+ \be v^i_2,\ar w^i_1+ \be w^i_2)$ in general define different points in $T^2_m\M$ for the same $\xi_1,\, \xi_2,\ \ar,\, \be$ if different initial coordinate charts on $\M$ were chosen. Surprisingly, one can find out by a direct verification that the expression $(x^i,\ar v^i,\ar^2w^i+\be v^i)$ really defines the same point in $T^2_m\M$ for all coordinate charts. Therefore, the expression in the equality \re{2repinvcond} is well-defined.

\begin{small}
\paragraph{Example.} The simplest non-degenerate example of \rit Lagrangian with the second derivatives is
\eq{
L = \dfrac{(\xtt\yt-\xt\ytt)^2}{(\xt^2+\yt^2)^{5/2}}.
}
It is a two-dimensional Euclidean version of the relativistic Lagrangian
\eq{
Ldt = \br{\dif{^2x^{\mu}}{s^2}}^2 ds = \dfrac{\xtt^{\mu}\xtt_{\mu}\xt^{\nu}\xt_{\nu}-\br{\xtt^{\mu}\xt_{\mu}}^2}{\br{\xt^{\la}\xt_{\la}}^{5/2}}dt,
}
where $s$ is the proper time.
This (term in the) Lagrangian is encountered in radiation theory (see \cite{M&M1,M&M2,Galakhov}). The Lagrangian containing this term, namely the Lagrangian for the relativistic particle with curvature, is also studied in, for example, \cite{Plush}.
\end{small}

\section{Construction for systems with the first derivatives}
\label{1case}
We start with constructing the Hamiltonian formalism in the simplest case of Lagrangians with the first derivatives.

Let us consider the Lagrangian system with the first derivatives. Denote by $x^i$, $i=1..n$, coordinates on $\M$. Then coordinates $(x^i,v^i)$ on $T\M$ and $(x^i,p_i)$ on $T^*\M$ are naturally induced.

Define the Legendre transformation $\leg:T\M \map T^*\M$, where $T\M$ is the tangent bundle to $\M$ and $T^*\M$ is the cotangent bundle, in the following conventional way:
\eql{ex}{
\leg: (x^i,v^i) \mapsto \br{x^i,\pdif{L}{v^i}}.
}
Note that $\pdif{L}{v^i}$ transform in the same way as $p_i$ under coordinate transformations. Therefore the map is well-defined.

Note that points in $T\M$ may be thought of as initial conditions for the Cauchy problem to the Euler-Lagrange equations. This means that given point in $T\M$ determines a unique solution of the Euler-Lagrange equation, i.e. a vector in some point of $\M$ defines a unique extremal of the action passing through this point. In the reparametrization invariant case it is obvious that all nonzero vectors which are proportional to each other with positive coefficient of proportionality define the same unparametrized oriented curve.

Denote the image of $\leg$ as $\PS$. It is called the \textit{phase bundle}. As it follows from equality \re{1repinvcond}
\eqn{L(m,\ar v)=\ar L(m,v),}
the map $\leg$ has the same value on each vector $(m,\ar v) \ \forall \ar > 0$, where $m\in \M,\; v\in T\M$. Therefore, $\dim\PS=2n-1$. Note that this one-parametric family $(m,\ar v)$ is exactly the above-mentioned set of points which define the same extremal curve.

Due to $\PS$ being a hyper-surface in $T^*\M$ (i.e. a $2n-1$-dimensional submanifold), it can be defined as a solution of the equation $\Phi(x,p) = 0$, where $\Phi(x^i,p_i)$ is some smooth function on $T^*\M$. $\Phi(x^i,p_i)$ is then called a \textit{constraint}. On $T^*\M$ as on every cotangent bundle there exists the canonical symplectic non-degenerate two-form $\om$. The restriction of $\om$ to $\PS$ defines a direction field on $\PS$ like every non-degenerate differential two-form on an odd-dimensional manifold (for more details see Section \ref{hf}). If $\PS$ is defined by the equation $\Phi = 0$ one can write the explicit formula for the integral curve of this field:
\eqs{
&\xt^i = c\pdif{\Phi}{p_i}, \\
&\pt_i = -c\pdif{\Phi}{x^i},
}
where c arbitrarily depends on the curve parameter. We call this system the \textit{Hamilton equations}. They can be rewritten in the following way:
\eqs{
&\xt^i = c\bd{x^i,\Phi} \\
&\pt_i = c\bd{p_i,\Phi},
}
where $\bd{\cdot,\cdot}$ is the usual Poisson bracket of functions on the cotangent bundle.

One can derive the Lagrange equations from the Hamilton equations. Note that $\displaystyle \Phi\br{x^i,\pdif{L}{v^i}(x^i,v^i)} = 0$ and $\displaystyle L=\pdif{L}{v^i}v^i$. Therefore,
\eq{
\dif{}{t}\pdif{L}{v^i} = \pt_i = -c\pdif{\Phi}{x^i} = c\pdifd{L}{v^j}{x^i}\pdif{\Phi}{p_j} = \pdifd{L}{v^j}{x^i}\xt^j = \pdifd{L}{v^j}{x^i}v^j = \pdif{L}{x^i},
}
i.e. the Lagrange equations are satisfied.

In Section \ref{gencon} we construct the Legendre transformation and the Hamilton field in the general case of higher derivatives. We considered above the construction of the Hamilton field in the case of the first derivatives. For better understanding we now consider the construction for the Legendre transformation in this simple case. That is, we obtain the conventional formula \re{ex}
\eqn{
\br{x^i,v^i} \mapsto \br{x^i,\pdif{L}{v^i}}
}
in the same way which will be used to obtain the generalization of this formula in the case of higher derivatives.

Consider a vector $\ve$ in some point $m_0 \in \M$. As was mentioned, it defines an extremal curve $\g$ passing through $m_0$. Fix a point $e$ on $\g$ not far from $m_0$. Let e be chosen on the opposite side of $\g$ to which $v$ points (the reason for doing so is shown below). Then in the neighbourhood of $m_0$ we define a function $\s$ in the following way: the value of $\s$ on some point $m$ is the value of the action on the unique extremal curve which connects $e$ and $m$. We use here the fact that for two given points on $\M$ there locally exist a unique extremal curve which connects them. Then we consider the differential of the function $\s$ in the point $m_0$. It is $d\s|_{m_0} \in T^*_{m_0}\M$. We define the Legendre transformation to be
\eq{
\leg : \br{m_0,\ve}\mapsto \br{m_0,d\s|_{m_0}}.
}
Here $\br{m_0,\ve}\in T\M$ and $\br{m_0,d\s|_{m_0}} \in T^*\M$.

To derive formulae in coordinates consider a variation of the action $S$ on the curve $\g$:
\eq{
\delta S = S(\g+h)-S(\g),
}
where $\g+h$ is a small deviation from curve $\g$, $h$ being a curve in the coordinate space $\RF^n$. Let us now introduce on $\g$ and $h$ some parametrization, so that they are functions $\g:[a,b]\map \M,\; h:[a,b]\map \RF^n$ and $\g(a)=e,\, \g(b) = m_0\ , h(a) =0$. The condition $h(a) = 0$ corresponds to varying the only $m_0$ end of $\g$. Now one can write
\eq{
\delta S = \intgd{a}{b}\br{L\br{\g^i+h^i,\gatr^i+\fht^i}-L\br{\g^i,\gatr^i}}dt.
}
Keeping the first order terms in $h,\, \fht$ we will have
\eq{
\delta S=
\intgd{a}{b}\br{h^i\pdif{L}{x^i}\br{\g^i\br{t},\gatr^i\br{t}} + \fht^i\pdif{L}{v^i}\br{\g^i\br{t},\gatr^i\br{t}}}dt,
}
where, as it was mentioned, $(x^i,v^i)$ are coordinates on $T\M$. Integrating the second term in the integrand by parts one obtains
\eqm{
\delta S=
\br{h^i\br{t}\pdif{L}{v^i}\br{\g^i\br{t},\gatr^i\br{t}}}\bigg|^b_a + \intgd{a}{b}\br{h^i\br{t}\pdif{L}{x^i}\br{\g^i\br{t},\gatr^i\br{t}} - h^i\br{t}\dif{}{t}\pdif{L}{v^i}\br{\g^i\br{t},\gatr^i\br{t}}}dt = \\
=\br{h^i\br{t}\pdif{L}{v^i}\br{\g^i\br{t},\gatr^i\br{t}}}\bigg|^b_a + \intgd{a}{b}\br{\pdif{L}{x^i}\br{\g^i\br{t},\gatr^i\br{t}} - \dif{}{t}\pdif{L}{v^i}\br{\g^i\br{t},\gatr^i\br{t}}}h^i\br{t}dt.
}
Recall now that the expression $\pdif{L}{x^i}\br{\g^i\br{t},\gatr^i\br{t}} - \dif{}{t}\pdif{L}{v^i}\br{\g^i\br{t},\gatr^i\br{t}}$, which has the form of the left hand side of the Euler-Lagrange equation, vanishes for all $t$ since $\g$ is an extremal of the action. Also note that, because $h(a) = 0$, only one boundary term contributes. Therefore we finally  have
\eq{
\delta S = h^i\br{b}\pdif{L}{v^i}\br{\g^i\br{b},\gatr^i\br{b}}.
}
Hence, due to the definition of $\s$ we have
\eq{
d\s|_{m_0} = \pdif{L}{v^i}\br{\g^i\br{b},\gatr^i\br{b}}dx^i.
}
As was mentioned, the functions $\pdif{L}{v^i}$ are constant on all points in $T_{m_0}\M$ defining the same extremal curve. The curve $\g$ was defined as an extremal curve corresponding to the vector $\ve$, that is in arbitrary parametrization the following holds:
\eq{
\gatr\br{b} = \ar \ve
}
for some positive $\ar$. $\ar$ is positive because e was chosen on the side of $\g$ opposite to which $\ve$ points. Therefore,
\eq{
\pdif{L}{v^i}\br{\g^i\br{b},\gatr^i\br{b}} = \pdif{L}{v^i}\br{x_0^i, v_0^i},
}
where $(x_0^i,v_0^i)$ are the coordinates of the point $(m_0,\ve)\in T\M$.
Hence one finally obtains
\eq{
d\s|_{m_0} = \pdif{L}{v^i}\br{x_0^i, v_0^i}dx^i.
}
Therefore, the formula for the Legendre transformation takes in coordinates the expected form \re{ex}:
\eqn{
\leg: (x^i,v^i) \mapsto \br{x^i,\pdif{L}{v^i}\br{x^i,v^i}}.
}

\begin{small}

\paragraph{Example.} Let us consider the above construction for the mentioned Lagrangian
\eq{
L = \sqrt{\xt^2+\yt^2}
}
on the plane $\RF^2 = \M$. We work in the notation introduced in this chapter.

In this example the formula for the differential of $\s$ can be obtained directly, without studying the variation of action. Let for simplicity the point $m_0$ be the origin. Let the vector $\ve$ have the coordinates $\br{v_{x0}, v_{y0}}$. The extremal corresponding to the vector $\ve$ is the straight line $\g$ in $\RF^2$ passing through the origin in the direction pointed by $\ve$. Let the point $e$ have coordinates $\br{x_0,y_0}$. The point $e$ should lie on the ray of $\g$ opposite to the direction of $\ve$. We have $\dfrac{x_0}{y_0} = \dfrac{v_{x0}}{v_{y0}}$ and the proportionality coefficient between $(x_0,y_0)$ and $\br{v_{x0}, v_{y0}}$ has a negative value. Because the action on the curve in this case is simply its length, and the extremals are just straight lines, the function $\s$ is written as
\eq{
\s\br{x,y} = \sqrt{\br{x-x_0}^2+\br{y-y_0}^2}.
}
Therefore its differential in the point $m_0$, the origin of coordinates, is expressed as
\eq{
d\s|_0 = -\dfrac{x_0}{\sqrt{x_0^2+y_0^2}}dx-\dfrac{y_0}{\sqrt{x_0^2+y_0^2}}dy = \dfrac{v_x}{\sqrt{v_x^2+v_y^2}}dx + \dfrac{v_y}{\sqrt{v_x^2+v_y^2}}dy = \pdif{L}{\xt}\br{v_x,v_y}dx+\pdif{L}{\yt}\br{v_x,v_y}dy.
}
Thus, the formula for the Legendre transformation takes the form
\eq{
\leg(x,y,v_x,v_y) = (x,y,\pdif{L}{\xt}\br{x,y,v_x,v_y},\pdif{L}{\yt}\br{x,y,v_x,v_y})
}
as was expected.

Due to the Legendre transformation being defined by
\eq{
\br{x,y,v_x,v_y} \mapsto \br{x,y,\dfrac{v_x}{\sqrt{v_x^2+v_y^2}},\dfrac{v_y}{\sqrt{v_x^2+v_y^2}}},
}
its image is defined by the equation $p_x^2+p_y^2=1$, i.e. the constraint takes the form $\Phi = p_x^2+p_y^2-1$. The Hamilton equations have the form
\eqs{
&\xt = 2 c p_x, \\
&\yt = 2 c p_y, \\
&\pt_x = 0, \\
&\pt_y = 0, \\
}
i.e. solutions of them are all possible lines on the plane, as was expected.

\end{small}

\section{General construction}
\label{gencon}
\subsection{Legendre transformation}
\label{genlegtrans}
In this section we consider the general case of the Lagrangian systems with $k$-th order derivatives. We are going to define the Legendre transformation, a map $\leg:T^{2k-1}\M \map T^*T^{k-1}\M$.

Let $m_2$ be a point in $T^{2k-1}\M$. Because for systems with $k$-th order derivatives the Euler-Lagrange equations are of order $2k$, the space $T^{2k-1}_m\M$ in some point $m \in \M$ can be understood as the space of the Cauchy data for the Euler-Lagrange equations in this point. Therefore there is a unique extremal curve $\g$ on $\M$ which can be lifted to $T^{2k-1}\M$ to pass through $m_2$ (``lifting'' a curve to the tangent bundle of some order means introducing a parametrization on that curve and then mapping it to its derivatives with respect to this parametrization).
Note that on every curve we have one distinguished parametrization -- \textit{parametrization by the action along that curve}\footnote{We implicitly assume here that the Lagrangian does not vanish on nonzero elements of $T^k\M$. In this case the action can always be used as a parameter. If it is not the case, it can be used as a parameter locally.}. This very fact will be the core one in the construction.

Consider this natural action parametrization on $\g$. It gives us a lift of $\g$ to $T^{k-1}\M$. Note that we now consider the space $T^{k-1}\M$, not $T^{2k-1}\M$. The latter is the space of the Cauchy data; whilst the former is the space of data for the boundary problem, that is if two close enough points in $T^{k-1}\M$ are fixed, there exists a unique extremal curve which can be lifted to connect them. Let $m_1$ be the point obtained by lifting $m_0$ to $T^{k-1}\M$ using the natural parametrization of $\g$, where $m_0\in \M$ is a projection of $m_2$ to $\M$. Let us arbitrarily choose point $e$ on $\tg$ close to $m_1$, where $\tg$ is the mentioned lift of $\g$. Let $e$ be chosen on the side of $\tg$ opposite to the side to which $m_2$ is pointing. This is needed for the tangent vector to $\g_{em_1}$ at $m_1$ to point to the same direction to which $m_2$ points.

We now define a function $\s$ on some small neighbourhood $U$ of $m_1$ in the following way. Let $r$ be a point in $U$. Then there exists a unique extremal curve which connects $e$ and $r$. So, let us define the value of function $\s$ in point $r$ as simply the value of the action on this curve. That is, $\s(r) = S(\g_{er})$, where $\g_{er}$ is this unique extremal curve which can be lifted to $T^{k-1}\M$ to end up in the points $e$ and $r$. Now having defined in such a way the function $\s$ on $U$, we can consider its differential at the point $m_1$, i.e. $p = d\s|_{m_1}$. Note that $p \in T^*_{m_1}T^{k-1}\M$.

 So, we define the Legendre transformation as follows:
\eq{
	\leg(m_2) := d\s|_{m_1}.
}

Note that all points in $T^{2k-1}\M$ which define one and the same extremal curve as the Cauchy data are mapped into one point by the Legendre transformation. Indeed, the only fact about the point in $T^{2k-1}\M$ which was used in the construction is the (oriented) extremal curve it defines.

Now we shall find the way to obtain explicit formulae for the Legendre transform. By the way, note that we have not shown so far that the Legendre transformation is well-defined. Indeed, it could in principle depend on the choice of $e$. However, the explicit formulae show that this is not the case and therefore the transformation \textit{is} well-defined.

Let us think of the above-mentioned curve $\g$ as a curve with ending points $e_0,\, m_0$ respectively, where $e_0$ is the projection of $e$ onto $\M$. For a while we consider an arbitrary parametrization on $\g$, i.e. $\g:[a,b]\map \M,\; \g(a)=e_0,\, \g(b)=m_0$. The action on $\g$ is expressed as
\eq{
S(\g) = \intgd{a}{b}L\big(C^k_{\g(t)}(\g)\big)dt,
}
where $C^k_{\g(t)}(\g) \in T^k_{\g(t)}\M$ is the class of equivalence of the curve $\g$ in the point $\g(t)$. In coordinates it is expressed as
\eq{
S(\g) = \intgd{a}{b}L\br{\g^i,\dot{\g}^i,\ddot{\g}^i,\dots,\gdi{k}}dt.
}

Consider in coordinates on $\M$ a small deviation from the curve $\g$ : the curve $\g+h$, where $h:[a,b]\map \RF^n$, $\displaystyle h(a) = 0,\,\dot{h}(a)=0,\,\dots,\,\mathop{h}^{\scriptscriptstyle(k-1)}=0$. Let $\br{x^i,\x{1},\dots,\x{2k-1}}$ be the coordinates on $\ttkm$. Let us write the variation of the action $S$:
\eqn{
\delta S = S(\g+h)-S(\g) =
\intgd{a}{b}\br{L\br{\g^i+h^i,\dot{\g}^i+\dot{h}^i,\ddot{\g}^i+\ddot{h}^i,\dots,\gdi{k}+\hdi{k}}-L\br{\g^i,\dot{\g}^i,\ddot{\g}^i,\dots,\gdi{k}}}dt =
}
\eqn{
= h^i(b)\br{\pdif{L}{x_{(1)}^i}-\dif{}{t}\pdif{L}{\x{2}}+\dots+(-1)^{k-1}\dif{^{k-1}}{t^{k-1}}\pdif{L}{\x{k}}}
+
}
\eqn{
+\dot{h}^i(b)\br{\pdif{L}{x_{(2)}^i}-\dif{}{t}\pdif{L}{\x{3}}+\dots+(-1)^{k-2}\dif{^{k-2}}{t^{k-2}}\pdif{L}{\x{k}}}+
}
\eqn{
+\dots+
}
\eql{genactvar}{
+\hdi{k-1}(b)\pdif{L}{\x{k}},
}
where the derivative $\dif{}{t}$ being applied to functions on $\ttkm$ is assumed to mean simply $\x{1}\pdif{}{x^i}+\x{2}\pdif{}{\x{1}}+\dots+\x{2k-1}\pdif{}{\x{2k-2}}$. All expressions on the rightmost side are assumed to be taken in the point $C^{2k-1}_{\g(b)}(\g)\in T^{2k-1}_{\g(b)}\M$: the class of equivalence of the arbitrary-parametrized curve $\g$ in the point $m_0$. For more details on the formula \re{genactvar} see Section \ref{variation} in the Appendix.

Now we return to parametrization of $\g$ by the action and from the definition of the function $\s$ obtain the expression for its differential in point $m_1$:
\eqn{
d\s|_{m_1} = \br{\pdif{L}{x_{(1)}^i}-\dif{}{t}\pdif{L}{\x{2}}+\dots+(-1)^{k-1}\dif{^{k-1}}{t^{k-1}}\pdif{L}{\x{k}}}dx
+}
\eqn{
+\br{\pdif{L}{x_{(2)}^i}-\dif{}{t}\pdif{L}{\x{3}}+\dots+(-1)^{k-2}\dif{^{k-2}}{t^{k-2}}\pdif{L}{\x{k}}}d\x{1}+
}
\eqn{
+\dots+}
\eq{
+\pdif{L}{\x{k}}d\x{k-1},
}
where now all the expressions are taken in the point $C^{2k-1}_{m_0}\br{\g}\in T^k_{m_0}\M$: the class of equivalence of the action-parametrized curve $\g$ in the point $m_0$.

We see that the expression for $d\s$ contains no dependence on the particular choice of $e$ and therefore the Legendre transformation is well-defined.

To express the formula of the Legendre transformation in terms of coordinates $\br{y^i,\y{1},\dots,\y{2k-1}}$ of the initial point $m_2\in T^{2k-1}\M$ we need to find the coordinates of $C^{2k-1}_{m_0}(\g)$ first. Because the natural action-parametrization of $\g$ was considered, they are
\eq{
\br{\g^i(b_0),\,\dif{\g^i}{s}(b_0),\,\dots,\,\dif{^k\g^i}{s^k}(b_0)},
}
where $b_0 = S\br{\g_{em_1}}$, i.e. $\g\br{b_0} = m_1$.
Now note that $\dif{\g}{s}=\dfrac{1}{L}\dif{\g}{t},$ where $L$ is taken at the corresponding point. Denote $L$ taken in this point as $L_0$, $\dif{L}{t}$ as $\dot{L}_0$, and so on. So, the coordinates of $C^{2k-1}_{m_0}(\g)$ become rewritten as
\eq{
\br{y^i, \dfrac{1}{L_0}\y{1},\,\dfrac{1}{L_0^2}\y{2}-\dfrac{\dot{L}_0}{L_0^3}\y{1},\,\dfrac{1}{L_0^3}\y{3}-3\dfrac{\dot{L}_0}{L_0^4}\y{2}+\br{3\dfrac{\dot{L}_0^2}{L_0^5}-\dfrac{\ddot{L}_0}{L_0^4}}\y{1},\,\dots}
}
All these coordinates are obtained by differentiation
\eq{
\dfrac{1}{L}\dif{}{t}\dfrac{1}{L}\dif{}{t}\dots\dfrac{1}{L}\dif{}{t}\g^i,
}
and then taking the result at the corresponding point.

Finally, the formula for the Legendre transformation is obtained by substituting this coordinates of the point $C^{2k-1}_{m_0}(\g)$ into the expression of $d\s|_{m_1}$. Using the relations of the kind \re{1repinvcond}, \re{2repinvcond} (higher order Zermelo conditions) for \rit Lagrangian with $k$-th derivatives, one can extract factors of $L_0,\,\dot{L}_0,\dots$ from the arguments of the functions. To make this more transparent, we describe below in detail the construction for the case of the second derivatives.

This construction can be thought of as being simply the generalization of the concepts of relativistic mechanics. The Legendre transformation maps a point in $\ttkm$ to some covector from $\tkm$. The coordinates of this point in $\tkm$ are analogous to the ``four-velocities'' of relativistic mechanics, as they are obtained as derivatives of the curve in the action parametrization, that is, the ``proper time''. The covector $d\s|_{m_1}$ is simply the generalization and more formal notation of the definition of momentum $\pdif{S}{x}$ used in relativistic mechanics (see \cite{LL}).

\subsection{Hamilton field}
\label{hf}
Because every $2k-1$-parametric family of points in $T^{2k-1}\M$ which define the same extremal curve is mapped entirely to one point, the dimension of the image of the Legendre transformation is reduced by the factor of $2k-1$ from the value $2kn$ of the dimension of $T^{2k-1}\M$. That is,
\eq{
\dim\big(\leg(T^{2k-1}\M)\big)=2kn-(2k-1),
}
which is an odd number. Therefore, the manifold $\PS = \leg(T^{2k-1}\M) \subset T^*\tkm$ is an odd-dimensional submanifold. Note that $\PS$ can be thought of as a bundle over $\tkm$, and also as a bundle over $\M$.

The bundle $T^*\tkm$ as any cotangent bundle has the canonical symplectic structure. The restriction $\om_{\PS}$ of the canonical symplectic two-form $\om$ to submanifold $\PS$ is therefore differential two-form on an odd-dimensional manifold. As every non-degenerate differential two-form on an odd-dimensional manifold, $\omp$ determines a direction in every point of $\PS$, i.e. some one-dimensional subspace of the tangent space (see Section \ref{nambu} for more details). This subspace $V_{\chi}$ in point $\chi \in \PS$ is defined as the space of all vectors $\eta\in T_{\chi}\PS$ which satisfy the condition
\eq{
\omp(\eta)=0.
}
Because $\omp$ is non-degenerate two-form on an odd-dimensional manifold, $V_{\chi}$ is exactly one-dimensional.

The collection of this spaces $V_{\chi}$ over all $\PS$ is called the \textit{Hamilton field}. It is a field of directions, not a vector field, which is in agreement with reparametrization invariance. Because the projections of integral curves of this field to $\M$ are desired to be extremals of the action, one is not interested in their parametrization, and the field with unparametrized integral curves is a field of directions.

We can write explicit formula for $V_{\chi}$ using the so-called \textit{generalized Nambu bracket} (see Section \ref{nambu}). If $\eta \in V_{\chi}$ is a nonzero vector, $\eta = \eta^{\mu}\pdif{}{\xi^{\mu}}$, where $\xi = (x,p)$ ($x^i,\, p^i$ are coordinates on $T^*\tkm$), then the following holds:
\eql{hamnambu}{
\eta^{\mu} = \bd{\xi^{\mu},\Phi_1,\dots,\Phi_{2k-1}},
}
where the bracket is the generalized Nambu bracket between $2k$ functions. See Section \ref{nambu} for derivation of the \re{hamnambu} and for more details on the subject.

Now let us prove that the problem of finding extremals of the action is transformed to the problem of finding integral curves, as was desired.

First we shall see that for every curve $\g$ in $\M$
\eq{
S(\g) = \intg{\gp}\theta_{\PS}.
}
Here $\gp$ is a curve in $\PS$. This curve is obtained by lifting $\g$ to $\ttkm$ via introduction of some parametrization, and then performing the Legendre transformation. Note that the image under the Legendre transformation does not depend on the choice of parametrization, as it follows from what we have discussed earlier. Recall that $\PS$ is a bundle over $T^{k-1}\M$ and note that the projection of $\gp$ to $T^{k-1}\M$ is the curve $\tg$, exactly the action-parametrized lift of $\g$ to $\tkm$. The differential one-form $\theta_{\PS}$ is the restrtiction of the canonical one-form $\theta$ from $T^*\tkm$ to $\PS$.

Note that by definition of the canonical one-form, in the point $(m_1,p)\in \PS$, where $m_1\in\tkm,\ p\in T^*_{m_1}\tkm$, the following relation holds:
\eq{
\theta_{_{\PS}(m_1,p)}(\eta) = p(\pi_{\PS}'(\eta)),
}
where $\eta\in T_{(m_1,p)}\PS$, and $\pi_{\PS}' \in \Hom(T_{(m_1,p)}\PS,T_{m_1}\tkm)$ is the derivative of the map $\pi_{\PS}$ of projection from $\PS$ to $\tkm$.

Let us introduce some parametrization on $\gp$ and, therefore, parametrization on $\tg$. Thus, one has $\gp:[a,b]\map \PS$ and $\tg:[a,b]\map \tkm$. Recall the definition of the integral of one-form over the curve:
\eq{
\intg{\gp}\theta_{\PS} = \intgd{a}{b}\theta_{\PS}(\gpt(t))dt,
}
where $\gpt(t)$ is the tangent vector to $\gp$ in the point $\gp(t)$. Note that $\pi_{\PS}'(\gpt(t))=\dot{\tg}(t)$ because $\tg$ is the projection of $\gp$.

Recall from the definition of the Legendre transformation that the point $\gp(t)$ is really the pair $\br{\tg(t),\, d\s|_{\tg(t)}}$ where $d\s|_{\tg(t)}\in T^*_{\tg(t)}\tkm$ is the differential of the function $\s$ described in the previous subsection. Note that because this differential does not depend on the choice of starting point $e$, without loss of generality we can take the point $e$ to be $\tg(a)$. Then all points of $\gp$ are obtained using the same function $\s$. Recalling the definition of the canonical one-form one then has
\eq{
\theta_{\PS}(\gpt(t)) = d\s|_{\tg(t)}\br{\dot{\tg}(t)}.
}
However the integral of the differential of some function along some curve is simply the difference in the values of that function in the endpoints:
\eq{
\intgd{a}{b}d\s|_{\tg(t)}\br{\dot{\tg}(t)}dt = \intg{\tg}d\s = \s\br{\tg(b)}-\s\br{\tg(a)}.
}
From our definition of $\s$ one has $\s\br{\tg(a)} = 0,\ \s\br{\tg(b)} = S\br{\tg}$. Therefore one finally obtains the desired equality
\eqn{
S(\g) = \intg{\gp}\theta_{\PS}.
}

So it follows that the extremals of the action integral on manifold $\M$ are projections of the extremals of integral of the restriction of the canonical one-form to the manifold $\PS$. Consider a deviation from the curve $\gp$, the curve $\gp'$. The increment of integral of the canonical one-form
\eq{
\intg{\gp'}\theta_{\PS} - \intg{\gp}\theta_{\PS}
}
is equal to the symplectic area of the surface connecting the curves $\gp,\, \gp'$ and therefore is an infinitesimal of order higher than the difference between $\gp$ and $\gp'$ in the case when $\gp$ is an integral line of null-directions of $\omp$ (it follows from the Stokes formula). For more details on symplectic geometry see, for example, \cite{ArGi}.

Hence, the integral lines of null-directions of $\omp$ are extremals of integral of $\theta$ and therefore project to extremals of the action on $\M$.

\section{Construction for systems with the second order Lagrangian}
\label{2case}

Let us consider the reparametrization invariant action with second-order derivatives
\eql{fun}{
S = \intgd{a}{b} L(\g^i,\dot{\g}^i,\ddot{\g}^i) dt.
}

The Euler-Lagrange equations are
\eq{
\pdif{L}{x^i}-\dfrac{d}{dt} \br{\pdif{L}{v^i}-\dfrac{d}{dt} \pdif{L}{w^i}}=0,
}
where $(x^i,v^i,w^i)$ are coordinates on $T^2\M$, $\M$ is a configuration manifold.

These are differential equations of order four, which solutions are extremals of functional \re{fun}. Let us fix some point $m \in \M$. To choose the only extremal curve passing through $m$, one needs to define values of the first, second and third derivatives at this point. That is, $T^3_m\M$ is the space of the initial conditions for the Cauchy problem at the point $m$ (the definition of $T^k_m\M$ see in Section \ref{HD}).

Let $m_2$ be a point in $T^3_m\M$, then this point uniquely determines the extremal curve passing through $m$. However for every point $\br{x_0,v_0,w_0,u_0}\in T^3\M$ there is a 3-parametric family of the initial conditions for the Cauchy problem, every point of which determines the same extremal:
\eql{semeistvo}{
(x_0^i, \ \ar v_0^i, \ \ar^2 w_0^i + \be v_0^i, \ \ar^3 u_0^i + 3\ar\be w_0^i + \g v_0^i),
}
where $\ar > 0, \; \be, \, \g$ are arbitrary parameters (it is explained in Appendix \ref{prilozhenie} why the family has exactly this form). Thereby, one wants to define some space $\PS_m$ over $m$. There should be, roughly speaking, a one-to-one correspondence between all the extremal curves passing through the point $m$ in $\M$ and all the points in $\PS_m$ projecting to $m$. In other words, we make a transition from $T^3_m\M$ to $\PS_m$. For this reason one constructs the map
\eq{
\leg: T^3\M \map T^*T\M,
}
where $T^3\M = \displaystyle \mathop{\cup}_{m \in \M}T^3_m\M, \ \PS = \displaystyle \mathop{\cup}_{m \in \M}\PS_m$, $\PS=\leg(T^3\M)$; $\leg$ is the Legendre transformation, $\PS$ is a phase bundle.

Let us fix a point $m_2 \in T^3\M$. Let $m_0$ be the projection of the point $m_2$ to $\M$, and $\g$ be an extremal curve on $\M$, corresponding to $m_2$ as the solution to the Euler-Lagrange equations with Cauchy data given by the point $m_2$. Note that on every curve we have one distinguished parametrization: \textit{parametrization by the action along that curve}. Then let $\tg$ be the lift of $\g$ to $T\M$ obtained with this action parametrization on $\g$. The point $m_0 \in \g$ lifts this way to some point $m_1\in T\M$. Note that the space $T\M$ can be understood as the space of data for the boundary problem to the Euler-Lagrange equations.

We now define a function $\s$ on some small neighbourhood $U$ of $m_1$ in the following way. Let $r$ be a point in $U$. Then there exists a unique extremal curve which connects $e$ and $r$. Thus, let us define the value of function $\s$ in point $r$ as simply the value of the action on this curve. That is, $\s(r) = S(\g_{er})$, where $\g_{er}$ is this unique extremal curve which can be lifted to $T\M$ to end up in the points $e$ and $r$. Now having defined in such a way the function $\s$ on $U$, one can consider its differential at the point $m_1$, i.e. $p = d\s|_{m_1}$. Note that $p \in T^*_{m_1}T\M$.

Thus, one defines the Legendre transformation as follows:
\eq{
	\leg(m_2) := d\s|_{m_1}.
}

All points of the family \re{semeistvo} are mapped to the only point. Indeed, all points of this family determine the same extremal curve, therefore, from the definition of the function $\s$ one obtains that the value of the differential $d\s$ does not depend on the points of the family.

Let us write down explicit formulae for the Legendre transformation.

For a while we consider an arbitrary parametrization on $\g$. One can think of the curve $\g$ as a curve with ending points $e_0$ and $m_0$ respectively, where $e_0$ is the projection of $e$ onto $\M$. Let $\g+h$ be a small deviation from $\g$, then we introduce some parametrization on $\g$ and $h$; $\g:[a,b] \map \M, \ \g(a)=e_0, \ \g(b)=m_0, \ h:[a,b] \map \RF^n, \ h(a)=0,\;\dot{h}(a)=0.$ Now let us consider the variation of the action $S$:
\eq{
\delta S = S(\g+h)-S(\g) = \br{\pdif{L}{v^i}(\g^i,\dot{\g}^i,\ddot{\g}^i) - \br{\dfrac{d}{dt}\pdif{L}{w^i}}(\g^i,\dot{\g}^i,\ddot{\g}^i)}h^i(b) + \pdif{L}{w^i}(\g^i,\dot{\g}^i,\ddot{\g}^i)\dot{h}^i(b),
}
The derivative $\dif{}{t}$ here means simply 
\eql{tder}{
\dif{}{t} = v^i\pdif{}{x^i}+w^i\pdif{}{v^i}+u^i\pdif{}{w^i}.
}

Hence, due to the definition of $\s$ one has
\eq{
d\s|_{m_1}=\br{\pdif{L}{v^i}\big(C^2_{m_0}(\g)) - \br{\dfrac{d}{dt}\pdif{L}{w^i}}\big(C^3_{m_0}(\g)\big)}dx^i + \pdif{L}{w^i}\big(C^2_{m_0}(\g)\big)dv^i,
}
where $C^2_{m_0}(\g)\in T^2_{m_0}\M,\;C^3_{m_0}(\g)\in T^3_{m_0}\M$ are classes of equivalence of the action-parametrized curve $\g$ in the point $m_0$. $\dfrac{d}{dt}\pdif{L}{w^i}$ depends on a point in $T^3_{m_0}\M$ because of the $u^i$ term entering the formula \re{tder}. We see that the expression for $d\s$ contains no dependence on the particular choice of $e$ and therefore the Legendre transformation is well-defined.

The coordinates of $C^3_{m_0}(\g)=C^3_{\g\br{b_0}}(\g)$, where $b_0 = S\br{\g}$ (i.e. $\g\br{b_0}=m_0$), are
\eq{
\br{\g^i(b_0),\dif{\g^i}{s}(b_0),\dif{^2\g^i}{s^2}(b_0),\dif{^3\g^i}{s^3}(b_0)},
}
where $s$ is the action parameter along $\g$, and the coordinates of $C^3_{m_0}(\g)$, are, correspondingly,
\eq{
\br{\g^i(b_0),\dif{\g^i}{s}(b_0),\dif{^2\g^i}{s^2}(b_0)}.
}
Therefore, $d\s$ takes the following form
\eqm{
d\s|_{m_1}=\br{\pdif{L}{v^i}\br{\g^i(b_0),\dif{\g^i}{s}(b_0),\dif{^2\g^i}{s^2}(b_0)} - \br{\dfrac{d}{dt}\pdif{L}{w^i}}\br{\g^i(b_0),\dif{\g^i}{s}(b_0),\dif{^2\g^i}{s^2}(b_0),\dif{^3\g^i}{s^3}(b_0)}}dx^i + \\ + \pdif{L}{w^i}\br{\g^i(b_0),\dif{\g^i}{s}(b_0),\dif{^2\g^i}{s^2}(b_0)}dv^i.
}

Now note that $\dif{\g}{s}=\dfrac{1}{L}\dif{\g}{t}=\dfrac{\dot{\g}}{L}$, where $L$ is taken at the corresponding point. Denote $L$ taken in this point as $L_0$, $\dif{L}{t}$ as $\dot{L}_0$, $\dif{^2L}{t^2}$ as $\ddot{L}_0$. Thus, the coordinates of $C^3_{m_0}(\g)$ become rewritten as
\eq{
\br{\g^i(b),\dfrac{\dot{\g}^i(b)}{L_0},\dfrac{\ddot{\g}^i(b)}{L_0^2}-\dfrac{\dot{L}_0}{L_0^3}\dot{\g}^i(b),\dfrac{1}{L_0^3}\dddot{\g}^i(b)-3\dfrac{\dot{L}_0}{L_0^4}\ddot{\g}^i(b)+\br{3\dfrac{\dot{L}_0^2}{L_0^5}-\dfrac{\ddot{L}_0}{L_0^4}}\dot{\g}^i(b)}.
}
Using the notation
\eqn{
\g^i(b)=x_0^i, \ \dot{\g}^i(b) = v_0^i, \ \ddot{\g}^i(b) =  w_0^i,\ \dddot{\g}^i(b) =  u_0^i,
}
we rewrite it as
\eq{
\br{x_0^i,\dfrac{v_0^i}{L_0},\dfrac{w_0^i}{L_0^2}-\dfrac{\dot{L}_0}{L_0^3}v_0^i,\dfrac{1}{L_0^3}u_0^i-3\dfrac{\dot{L}_0}{L_0^4}w_0^i+\br{3\dfrac{\dot{L}_0^2}{L_0^5}-\dfrac{\ddot{L}_0}{L_0^4}}v_0^i}.
}
Finally, the formula for the Legendre transformation is obtained by substituting of this coordinates of the point $C_2\br{\g(b)}$ into the expression of $d\s|_{m_1}$ and $d\s$ become rewritten as
\eqm{
d\s|_{m_1}=\Bigg(\pdif{L}{v^i} \br{x_0^i,\dfrac{v_0^i}{L_0},\dfrac{w_0^i}{L_0^2}-\dfrac{\dot{L}_0}{L_0^3}v_0^i} - \\ \br{\dfrac{d}{dt}\pdif{L}{w^i}} \br{x_0^i,\dfrac{v_0^i}{L_0},\dfrac{w_0^i}{L_0^2}-\dfrac{\dot{L}_0}{L_0^3}v_0^i,\dfrac{1}{L_0^3}u_0^i-3\dfrac{\dot{L}_0}{L_0^4}w_0^i+\br{3\dfrac{\dot{L}_0^2}{L_0^5}-\dfrac{\ddot{L}_0}{L_0^4}}v_0^i} \Bigg)dx^i + \\ + \pdif{L}{w^i} \br{x_0^i,\dfrac{v_0^i}{L_0},\dfrac{w_0^i}{L_0^2}-\dfrac{\dot{L}_0}{L_0^3}v_0^i} dv^i.
}

Let us recall relation \re{2repinvcond}:
\eqn{
L\br{x^i,\ar v^i,\ar^2w^i+\be v^i} = \ar L\br{x^i,v^i,w^i}.
}
Differentiating this relation with respect to $w^i$ one obtains
\eq{
\ar^2 \pdif{L}{w^i}\br{x^i,\ar v^i,\ar^2w^i+\be v^i} = \ar \pdif{L}{w^i}\br{x^i,v^i,w^i},
}
that is
\eq{
\pdif{L}{w^i}\br{x^i,\ar v^i,\ar^2w^i+\be v^i} = \dfrac{1}{\ar} \pdif{L}{w^i}\br{x^i,v^i,w^i}.
}
In the same way one can obtain
\eqm{
\pdif{L}{v^i}\br{x^i,\ar v^i,\ar^2w^i+\be v^i} - \br{\dfrac{d}{dt}\pdif{L}{w^i}}\br{x^i,\ar v^i,\ar^2w^i+\be v^i,\ar^3u^i+3\ar\be w^i+\epsilon v^i}= \\ =\pdif{L}{v^i}\br{x^i,v^i,w^i} - \br{\dfrac{d}{dt}\pdif{L}{w^i}}\br{x^i,v^i,w^i,u^i}.
}

Using the following notation:
\eqn{
\ar = \dfrac{1}{L_0}, \ \be=-\dfrac{\dot{L}_0}{L_0^3},
}
one obtains
\eq{
\pdif{L}{w^i}\br{x_0^i,\ar v_0^i,\ar^2w_0^i+\be v_0^i} = \dfrac{1}{\ar} \pdif{L}{w^i}\br{x_0^i,v_0^i,w_0^i},
}
\eq{
\br{\pdif{L}{v^i} - \dfrac{d}{dt}\pdif{L}{w^i}}\br{x_0^i,\ar v_0^i,\ar^2w_0^i+\be v_0^i,\ar^3u_0^i+3\ar\be w_0^i+\epsilon v_0^i}=\br{\pdif{L}{v^i} - \dfrac{d}{dt}\pdif{L}{w^i}}\br{x_0^i,v_0^i,w_0^i,u_0^i}
}
for arbitrary $\epsilon$.
Hence, the expression for $d\s$ takes the form
\eq{
d\s|_{m_1}=\br{\pdif{L}{v^i} \br{x_0^i,v_0^i,w_0^i} - \br{\dfrac{d}{dt}\pdif{L}{w^i}} \br{x_0^i,v_0^i,w_0^i,u_0^i} }dx^i  + L_0 \pdif{L}{w^i} \br{x_0^i,v_0^i,w_0^i} dv^i.
}

Therefore, formula for the Legendre transformation takes in coordinates the following form:
\eql{leg}{
\leg: \ \br{x^i, \ v^i, \ w^i, \ u^i} \mapsto \br{x^i, \ \dfrac{v^i}{L}, \ \dfrac{\partial L}{\partial v^i} \ -  \ \dfrac{d}{dt} \dfrac{\partial L}{\partial w^i}, \ L\pdif{L}{w^i}},
}
where all the expressions on the right side are taken in the point $(x^i, v^i, w^i, u^i)$.

Due to dimension of $\PS = \leg\br{T^3\M}$ being equal to $4n-3$, one can define $\PS$ by three equations
\eql{gam1}{
\left\{
\begin{array}{l}
 \\
\Phi_1(x^i,a^i,p_i,s_i)=0 ,\\[4mm]
\Phi_2(x^i,a^i,p_i,s_i)=0 ,\\[4mm]
\Phi_3(x^i,a^i,p_i,s_i)=0 ,\\[4mm]
\end{array}
\right.
}
where $\br{x^i,a^i,p_i,s_i}$ are the coordinates on $T^*T\M$, for some functions $\Phi_1,\Phi_2,\Phi_3$. There exists the symplectic non-degenerate canonically defined 2-form
\eq{
\om=dp_i \wedge dx^i + ds_i \wedge da^i
}
on $T^*T^1\M$.

For the vector $\xt^i\pdif{}{x^i}+\dot{a}^i\pdif{}{a^i}+\pt^i\pdif{}{p^i}+\dot{s}^i\pdif{}{s^i} = \dot{\xi}^{\mu}\pdif{}{\xi^{\mu}}$ from the null-direction of $\om$ on $\PS$ one has
\eq{
\dot{\xi}^{\mu} = c\bd{\xi^{\mu},\Phi_1,\Phi_2,\Phi_3},
}
where $\bd{\cdot,\cdot,\cdot,\cdot}$ is the 4-fold generalized Nambu bracket and $c$ is a constant (see Section \ref{nambu}).

Explicitly, the Hamilton equations therefore have the form
\eq{
\dot{x}^i= c\br{\bd{x^i,\Phi_1}\bd{\Phi_2,\Phi_3}+\bd{x^i,\Phi_2}\bd{\Phi_3,\Phi_1}+ \bd{x^i,\Phi_3}\bd{\Phi_1,\Phi_2} } ,
}
\eq{
\dot{a}^i= c\br{\bd{a^i,\Phi_1}\bd{\Phi_2,\Phi_3} +\bd{a^i,\Phi_2}\bd{\Phi_3,\Phi_1} + \bd{a^i,\Phi_3}\bd{\Phi_1,\Phi_2}},
}
\eq{
\dot{p}_i= c\big(\bd{p_i,\Phi_1}\bd{\Phi_2,\Phi_3}+\bd{p_i,\Phi_2}\bd{\Phi_3,\Phi_1}+ \bd{p_i,\Phi_3}\bd{\Phi_1,\Phi_2}\big),
}
\eq{
\dot{s}_i= c\big(\bd{s_i,\Phi_1}\bd{\Phi_2,\Phi_3}+\bd{s_i,\Phi_2}\bd{\Phi_3,\Phi_1}+ \bd{s_i,\Phi_3}\bd{\Phi_1,\Phi_2}\big),
}
where $\bd{\cdot,\cdot}$ is simply the Poisson bracket.

\begin{small}

\paragraph{Example.}
Let us consider the Lagrangian
\eq{
L=\dfrac{(\xtt \yt-\xt \ytt)^2}{(\xt^2 + \yt^2)^{5/2}}.
}

The Legendre transformation has the form:
\eqn{
\leg:\br{x,y,v_x,v_y,w_x,w_y,u_x,u_y} \mapsto \br{x,y,a_x,a_y,p_x,p_y,s_x,s_y},
}
where
\eqn{
a_x = \dfrac{v_x(v_x^2+v_y^2)^{5/2}}{(w_x v_y-v_x w_y)^2},
}
\eqn{
a_y = \dfrac{v_y(v_x^2+v_y^2)^{5/2}}{(w_x v_y-v_x w_y)^2},
}
\eqn{
p_x = \dfrac{-4 w_x v_y v_x^2 w_y + 6 w_x v_y^3 w_y +5 w_x^2 v_y^2 v_x - v_x^3 w_y^2 - 6 v_x w_y^2 v_y^2 -2 v_y^2 u_x v_x^2-2 v_y^4 u_x +2 v_y v_x^3 u_y+2 v_y^3 v_x u_y}{(v_x^2+v_y^2)^{7/2}},
}
\eqn{
p_y = \dfrac{-(6 w_x^2 v_y v_x^2+w_x^2 v_y^3+4 w_x v_y^2 v_x w_y-6 v_x^3 w_y w_x-5 v_x^2 w_y^2 v_y-2 v_x^3 u_x v_y-2 v_x u_x v_y^3+2 v_x^4 u_y+2 v_x^2 u_y v_y^2)}{(v_x^2+v_y^2)^{7/2}},
}
\eqn{
s_x = -2\,{\frac { \left( {\it v_x}\,{\it w_y}-{\it v_y}\,{\it w_x} \right) ^{3}{\it v_y}}{ \left( {{\it v_x}}^{2}+{{\it v_y}}^{2} \right) ^{5}}},
}
\eqn{
s_y = 2\,{\frac { \left( {\it v_x}\,{\it w_y}-{\it v_y}\,{\it w_x} \right) ^{3}{\it v_x}}{ \left( {{\it v_x}}^{2}+{{\it v_y}}^{2} \right) ^{5}}}.
}

Therefore $\PS$ can be determined by the system of the 3 equations
\eq{
\left\{
\begin{array}{l}
 \\
\ s_x v_x + s_y v_y = 0 ,\\[4mm]
\ p_x v_x + p_y v_y - 1 = 0 ,\\[4mm]
\ \br{v_x^2+v_y^2}^5 - 16 \dfrac{v_x^2 v_y^2}{s_x^2 s_y^2} = 0 .\\[4mm]
\end{array}
\right.
}

Let us denote $s_x v_x + s_y v_y$ by $\Phi_1$, $p_x v_x + p_y v_y - 1$ by $\Phi_2$, $\br{v_x^2+v_y^2}^5 - 16 \dfrac{v_x^2 v_y^2}{s_x^2 s_y^2}$ by $\Phi_3$. In these terms the Hamilton field is determined by the Nambu bracket
\eq{
\dot{\xi}^{\mu} = c\bd{\xi^{\mu},\Phi_1,\Phi_2,\Phi_3},
}
where $\dot{\xi}^{\mu}$ stands for one of the Hamilton field components. That is, the Hamilton equations take the following form:

\eqn{
\xt =a_x \left(  \left( 10\, \left( {a_x}^{2}+{a_y}^{2} \right) ^{4}a_x-32\,{\frac {a_x{a_y}^{2}}{{{\it s_x}}^{2}{{\it s_y}}^{2}}} \right) a_x+ \left( 10\, \left( {a_x}^{2}+{a_y}^{2} \right) ^{4}a_y-32\,{\frac {{a_x}^{2}a_y}{{{\it s_x}}^{2}{{\it s_y}}^{2}}} \right) a_y \mbox{}-64\,{\frac {{a_x}^{2}{a_y}^{2}}{{{\it s_x}}^{2}{{\it s_y}}^{2}}} \right),
}
\eqn{
\yt =a_y \left(  \left( 10\, \left( {a_x}^{2}+{a_y}^{2} \right) ^{4}a_x-32\,{\frac {a_x{a_y}^{2}}{{{\it s_x}}^{2}{{\it s_y}}^{2}}} \right) a_x+ \left( 10\, \left( {a_x}^{2}+{a_y}^{2} \right) ^{4}a_y-32\,{\frac {{a_x}^{2}a_y}{{{\it s_x}}^{2}{{\it s_y}}^{2}}} \right) a_y \mbox{}-64\,{\frac {{a_x}^{2}{a_y}^{2}}{{{\it s_x}}^{2}{{\it s_y}}^{2}}} \right),
}
\eqn{
\dot{a}_x = a_x \left( 32\,{\frac {{\it p_x}\,{a_x}^{2}{a_y}^{2}}{{{\it s_x}}^{3}{{\it s_y}}^{2}}}+32\,{\frac {{\it p_y}\,{a_x}^{2}{a_y}^{2}}{{{\it s_x}}^{2}{{\it s_y}}^{3}}} \mbox{} \right) +32\,{\frac {{a_x}^{2}{a_y}^{2} \left( -{\it p_x}\,a_x-{\it p_y}\,a_y \right) }{{{\it s_x}}^{3}{{\it s_y}}^{2}}},
}
\eqn{
\dot{a}_y = a_y \left( 32\,{\frac {{\it p_x}\,{a_x}^{2}{a_y}^{2}}{{{\it s_x}}^{3}{{\it s_y}}^{2}}}+32\,{\frac {{\it p_y}\,{a_x}^{2}{a_y}^{2}}{{{\it s_x}}^{2}{{\it s_y}}^{3}}} \mbox{} \right) +32\,{\frac {{a_x}^{2}{a_y}^{2} \left( -{\it p_x}\,a_x-{\it p_y}\,a_y \right) }{{{\it s_x}}^{2}{{\it s_y}}^{3}}},
}
\eqn{
\pxt = 0,
}
\eqn{
\pyt = 0,
}
\eqmn{
\dot{s}_x = -{\it s_x}\, \left( 32\,{\frac {{\it p_x}\,{a_x}^{2}{a_y}^{2}}{{{\it s_x}}^{3}{{\it s_y}}^{2}}}+32\,{\frac {{\it p_y}\,{a_x}^{2}{a_y}^{2}}{{{\it s_x}}^{2}{{\it s_y}}^{3}}} \right) \mbox{}-\\
-{\it p_x}\, \left(  \left( 10\, \left( {a_x}^{2}+{a_y}^{2} \right) ^{4}a_x-32\,{\frac {a_x{a_y}^{2}}{{{\it s_x}}^{2}{{\it s_y}}^{2}}} \right) a_x+ \left( 10\, \left( {a_x}^{2}+{a_y}^{2} \right) ^{4}a_y-32\,{\frac {{a_x}^{2}a_y}{{{\it s_x}}^{2}{{\it s_y}}^{2}}} \right) a_y \mbox{}-64\,{\frac {{a_x}^{2}{a_y}^{2}}{{{\it s_x}}^{2}{{\it s_y}}^{2}}} \right)+\\
+ \left( -10\, \left( {a_x}^{2}+{a_y}^{2} \right) ^{4}a_x+32\,{\frac {a_x{a_y}^{2}}{{{\it s_x}}^{2}{{\it s_y}}^{2}}} \right)  \left( -{\it p_x}\,a_x-{\it p_y}\,a_y \right),
}
\eqmn{
\dot{s}_y = -{\it s_y}\, \left( 32\,{\frac {{\it p_x}\,{a_x}^{2}{a_y}^{2}}{{{\it s_x}}^{3}{{\it s_y}}^{2}}}+32\,{\frac {{\it p_y}\,{a_x}^{2}{a_y}^{2}}{{{\it s_x}}^{2}{{\it s_y}}^{3}}} \right) \mbox{}-\\
-{\it p_y}\, \left(  \left( 10\, \left( {a_x}^{2}+{a_y}^{2} \right) ^{4}a_x-32\,{\frac {a_x{a_y}^{2}}{{{\it s_x}}^{2}{{\it s_y}}^{2}}} \right) a_x+ \left( 10\, \left( {a_x}^{2}+{a_y}^{2} \right) ^{4}a_y-32\,{\frac {{a_x}^{2}a_y}{{{\it s_x}}^{2}{{\it s_y}}^{2}}} \right) b\mbox{}-64\,{\frac {{a_x}^{2}{a_y}^{2}}{{{\it s_x}}^{2}{{\it s_y}}^{2}}} \right)+ \\
+ \left( -10\, \left( {a_x}^{2}+{a_y}^{2} \right) ^{4}a_y+32\,{\frac {{a_x}^{2}a_y}{{{\it s_x}}^{2}{{\it s_y}}^{2}}} \right)  \left( -{\it p_x}\,a_x-{\it p_y}\,a_y \right).
}

\end{small}

\section{Symplectic form null-vectors and the generalized Nambu bracket}
\label{nambu}

In this section we show that the null-vectors of the restriction of the symplectic form to an odd-dimensional submanifold are determined by the generalized Nambu bracket (for the definition of it see \cite{Takhtajan, CurtZach}).

Consider a $2n$-dimensional symplectic manifold $\N$ with symplectic two-form $\om$ (in this paper it is used for $\N = T^*\tkm$ with canonically defined $\om$). Let $\PS \subset \N$ be a $2n-(2k-1)$-dimensional submanifold, for some $k \leq n$ (in this paper, again, the results of this section are used in the case of $\PS$ being the phase bundle, i.e. the image of the Legendre transformation).

Let us see that the restriction $\omp$ of $\om$ to $\PS$ defines in every point of $\PS$ a distinguished direction, i.e. a one-dimensional subspace of the tangent space. Because $\om$ is a non-degenerate differential two-form then $\omp$ is also non-degenerate. However, a non-degenerate skew-symmetric two-form on an odd-dimensional space has the canonical form
\eq{
\left( \begin{array}{ccc}
0 & -E & 0 \\
E & 0 & 0 \\
0 & 0 & 0 \end{array} \right),
}
where $E$ is an $(n-k)\times(n-k)$ identity matrix, if the dimension of $\PS$ is $2n-(2k-1)$.
Therefore, in every point $m$ of $\PS$ there is a distinguished one-dimensional subspace of $T_m\PS$ where $\omp$ vanishes.

Because $\PS$ is a $2n-(2k-1)$-dimensional submanifold of $\N$ it can be defined through a system of equations
\eq{
\left\{
\begin{aligned}
&\Phi_1 = 0, \\
&\Phi_2 = 0, \\
&\dots \\
&\Phi_{2k-1} = 0,
\end{aligned}
\right.
}
for some $2k-1$ smooth functions $\Phi_1,\dots,\Phi_{2k-1}$.

Note that the symplectic two-form $\om$ defines a correspondence between $T\N$ and $T^*\N$ via $\xi \mapsto \om(\xi)$ for $\xi \in T\N$. Because $\om$ is non-degenerate it is a one-to-one correspondence. Therefore we can map vectors from $T\N$ and their tensorial powers to covectors in $T^*\N$ and their tensorial powers correspondingly, and vice versa.

Consider the $k$-th exterior power of the symplectic two-form $\om$: the $2k$-form $\om^{\wedge k} \in \Omega^{2k}\N = \Lambda^{2k}T^*\N$. Its image in $\Lambda^{2k}T\N$ via the above-mentioned map is given by
\eq{
 \pi_k\br{c_1,\dots,c_{2k}} \defeq \om^{\wedge k}\big(\pi\br{c_1},\dots,\pi\br{c_{2k}}\big),
}
where $c_i$ are arbitrary covectors, and $\pi \in \Lambda^2T\N$ is the inverse of $\om$.

Now we will show that the vector $\pi_k\br{d\Phi_1,\dots,d\Phi_{2k-1}}$ taken in some point $m\in\PS$ spans the mentioned one-dimensional distinguished subspace. It is really a vector because if one substitutes $2k-1$ covectors into an element of $\Lambda^{2k}T\N$, one obtains a vector, i.e. an element of $T\N$.  Due to $\om$ being non-degenerate and therefore $\pi_k$ being non-degenerate and $d\Phi_i$ being independent, it is nonzero. How can we check that $\pi_k\br{d\Phi_1,\dots,d\Phi_{2k-1}}$ is really the desired vector? First, it shall be tangent to $\PS$, that is functions $d\Phi_i$ for all $i$ shall vanish on it. It is clear, because
\eq{
d\Phi_i\big(\pi_k\br{d\Phi_1,\dots,d\Phi_{2k-1}}\big) = \pi_k\br{d\Phi_1,\dots,d\Phi_{2k-1},d\Phi_i} = 0
}
due to skew-symmetricity of $\pi_k$. Second, the restriction $\omp$ of $\om$ to $\PS$ shall vanish on it, that is the result of applying $\om$ to it shall be a linear combination of $d\Phi_i$. Let $\xi$ be a vector. Then
\eqn{
\om\br{\pi_k\br{d\Phi_1,\dots,d\Phi_{2k-1}},\xi} = \pi_k\big(d\Phi_1,\dots,d\Phi_{2k-1},\om\br{\xi}\big) = \om^{\wedge k}\Big(\pi\big(d\Phi_1\big),\dots,\pi\big(d\Phi_{2k-1}\big),\pi\big(\om\br{\xi}\big)\Big) =
}
\eq{
= \om^{\wedge k}\br{\pi\br{d\Phi_1},\dots,\pi\br{d\Phi_{2k-1}},\xi}.
}
The last expression is the sum of terms of the form $\om\br{\pi\br{d\Phi_i},\xi}$ with some coefficients independent on $\xi$. Note that $\om\big(\pi\br{d\Phi_i},\xi\big) = \om\big(\pi\br{d\Phi_i}\big)\br{\xi} = d\Phi_i\br{\xi}$. Hence $\om\big(\pi_k\br{d\Phi_1,\dots,d\Phi_{2k-1}}\big)$ is a linear combination of $d\Phi_i$.

The tensor $\pi_k$ defines a $2k$-fold bracket on functions on $\N$ in the following way:
\eq{
\bd{f_1,\dots,f_{2k}} \defeq \pi_k\br{df_1,\dots,df_{2k}}.
}
Because $\pi_k$ was obtained from $\om^{\wedge k}$ we have for the bracket
\eq{
\bd{f_1,\dots,f_{2k}} = \summ{\mathrm{permutations} \ \s \ \mathrm{of} \ (1..2k)}{} \dfrac{\sign\br{\s}}{2^nn!}\big\{f_{\s_1},f_{\s_2}\big\}\dots\big\{f_{\s_{2k-1}},f_{\s_{2k}}\big\},
}
where $\bd{\cdot,\cdot}$ is the standard Poisson bracket induced by $\om$. This exactly coincides with the definition of the sub-maximal generalized Nambu bracket (see formula (8) of Section 2.2 in \cite{CurtZach}).

Therefore, if the distinguished vector has the form $\dot{\zeta_{\mu}}\pdif{}{\zeta_{\mu}}$ for coordinates $\zeta_{\mu}$ on $\N$, then its components are expressed as
\eq{
\dot{\zeta_{\mu}} = c\bd{\zeta_{\mu},\Phi_1,\dots,\Phi_{2k-1}},
}
where $c$ is a constant, common to all $\dot{\zeta_{\mu}}$, and $\{\displaystyle\mathop{\underbrace{\cdot,\dots,\cdot}}_{2k}\}$ is the $2k$-fold generalized Nambu bracket.

\section{Conclusion}

In this paper the Hamiltonian formalism for \rit systems with k-th order derivatives is constructed. The main point is that we choose a distinguished parametrization: the parametrization by the action along the curve. If one uses this fact, a well-defined Legendre transformation $\leg  : T^{2k-1}\M \rightarrow T^*T^{k-1}\M $ can be constructed. Its image is some submanifold $\PS \subset T^*T^{k-1}\M$ (phase bundle). It happens to always be an odd-dimensional manifold, from which fact arises a direction field (the Hamilton field) defined on $\PS$. Integral curves of this field are projected into extremal curves on the configuration manifold. For every fixed $k$ one can write the Hamilton equations which are equivalent to the Lagrange equations.

Note that it has a physical interpretation. Legendre transformation is a transition from the formalism of "coordinate-velocity" to the formalism of "coordinate-momentum". Our coordinate and momentum formulae resemble the  relativistic formulae for coordinate and momentum. It happens due to relativistic mechanics being a reparametrization invariant theory.

However many questions remain to be answered. How can one use achievements of the Nambu mechanics in this formalism? What does quantization mean in the terms of this formalism? How formalism changes if one maps surfaces instead of curves into configuration manifold? How does it connect with field theory and string theory? It would be interesting to investigate these topics.

\section{Acknowledgements}

The authors are grateful to Valery Dolotin, Andrei Mironov, Alexei Morozov, Alexander Popolitov and Vladimir Rubtsov for fruitful discussions and very helpful remarks. We also would like to thank all participants of Valery Dolotin's seminar and everybody, who discussed this work with us. This work is partly supported by the Russian  President's Grant of Support for the Scientific Schools NSh-8004.2006.2, by RFBR grant 07-02-00645 (P. Dunin-Barkowski) and by RFBR grant 07-02-00878 (A. Sleptsov).

\appendix

\section{Appendix}
\subsection{Three-parametric family of points in $T^3\M$ corresponding to the same extremal curve}
\label{prilozhenie}

Let us replace the curve parameter $ t $ by $f(t)$, where $f$ is an arbitrary differentiable function, then
$$ x \mapsto x $$
$$ \dot{x} \mapsto \dfrac{1}{\dot{f}(t)} \dot{x} $$
$$ \ddot{x} \mapsto \dfrac{1}{{\dot{f}(t)}^2} \ddot{x} - \dfrac{\ddot{f}(t)}{{\dot{f}(t)}^3} \dot{x} $$
$$ \dddot{x} \mapsto \dfrac{1}{{\dot{f}(t)}^3} \dddot{x} - 3 \dfrac{\ddot{f}(t)}{{\dot{f}(t)}^4} \ddot{x} + \br{3 \dfrac{{\ddot{f}(t)}^2}{{\dot{f}(t)}^5} - \dfrac{\dddot{f}(t)}{{\dot{f}(t)}^4}} \dot{x}. $$

Since $f$ is an arbitrary differentiable function, one can denote $$\ar := \dfrac{1}{\dot{f}(t_0)}, \ \be := - \dfrac{\ddot{f}(t_0)}{{\dot{f}(t_0)}^3}, \ \g := \ 3 \dfrac{{\ddot{f}(t_0)}^2}{{\dot{f}(t_0)}^5} - \dfrac{\dddot{f}(t_0)}{{\dot{f}(t_0)}^4},$$
where $t_0$ is the value of the parameter $t$ at the $x_0$. Therefore, replacing $t \mapsto f(t)$ one obtains
\eq{
(x^i,\ \dot{x}^i,\ \ddot{x}^i,\ \dddot{x}^i) \mapsto (x^i,\ \ar \dot{x}^i,\ \ar^2 \ddot{x}^i +\be \dot{x}^i,\ \ar^3\dddot{x}^i +3\ar \be \ddot{x}^i+\g \dot{x}^i).
}
Thus, if one can arbitrarily change parameter $t$ one has the 3-parametric family.

\subsection{Variation of the action}
\label{variation}

Here we derive the formula for the variation of the action which is used in Section \ref{genlegtrans}. We work in the notation introduced there.

\eq{
\delta S = S(\g+h)-S(\g) =
\intgd{a}{b}\br{L\br{\g^i+h^i,\dot{\g}^i+\dot{h}^i,\ddot{\g}^i+\ddot{h}^i,\dots,\gdi{k}+\hdi{k}}-L\br{\g^i,\dot{\g}^i,\ddot{\g}^i,\dots,\gdi{k}}}dt.
}
Keeping only first order terms in $h$ and its derivatives one obtains
\eq{
\delta S = \intgd{a}{b}\br{h^i\pdif{L}{x^i}\br{\g^i,\dots,\gdi{k}}+\dot{h}^i\pdif{L}{\x{1}}\br{\g^i,\dots,\gdi{k}}+\dots+\hdi{k}\pdif{L}{\x{k}}\br{\g^i,\dots,\gdi{k}}}dt.
}
Integrating all terms in the integrand but the first by parts we obtain
\eqm{
\delta S =
\Bigg(h^i\br{t}\pdif{L}{\x{1}}\br{\g^i\br{t},\dots,\gdi{k}\br{t}}+\dot{h}^i\br{t}\pdif{L}{\x{2}}\br{\g^i\br{t},\dots,\gdi{k}\br{t}}+\dots+\\
+\hdi{k-1}(t)\pdif{L}{\x{k}}\br{\g^i\br{t},\dots,\gdi{k}\br{t}}\Bigg)\Bigg|^b_a +\\
+\intgd{a}{b}\br{h^i\pdif{L}{x^i}\br{\g^i,\dots,\gdi{k}}-h^i\dif{}{t}\br{\pdif{L}{\x{1}}\br{\g^i,\dots,\gdi{k}}+\dots+\hdi{k}\pdif{L}{\x{k}}\br{\g^i,\dots,\gdi{k}}}}dt.
}
Note that because $\displaystyle h(a) = 0,\,\dot{h}(a)=0,\,\dots,\,\mathop{h}^{\scriptscriptstyle(k-1)}=0$, half of the boundary terms vanishes. That is, the expression may be rewritten as
\eqm{
\delta S =
h^i\br{b}\pdif{L}{\x{1}}\br{\g^i\br{b},\dots,\gdi{k}\br{b}}+\dot{h}^i\br{b}\pdif{L}{\x{2}}\br{\g^i\br{b},\dots,\gdi{k}\br{b}}+\dots+\\
+\hdi{k-1}(b)\pdif{L}{\x{k}}\br{\g^i\br{b},\dots,\gdi{k}\br{b}} +\\
+\intgd{a}{b}\br{h^i\pdif{L}{x^i}\br{\g^i,\dots,\gdi{k}}-h^i\dif{}{t}\br{\pdif{L}{\x{1}}\br{\g^i,\dots,\gdi{k}}+\dots+\hdi{k}\pdif{L}{\x{k}}\br{\g^i,\dots,\gdi{k}}}}dt.
}

Repeating this integration by parts starting from the third term in the integrand and so on, one finally obtains
\eqlm{varfin}{
\delta S = h^i(b)\br{\pdif{L}{x_{(1)}^i}-\dif{}{t}\pdif{L}{\x{2}}+\dots+(-1)^{k-1}\dif{^{k-1}}{t^{k-1}}\pdif{L}{\x{k}}}
+\\
+\dot{h}^i(b)\br{\pdif{L}{x_{(2)}^i}-\dif{}{t}\pdif{L}{\x{3}}+\dots+(-1)^{k-2}\dif{^{k-2}}{t^{k-2}}\pdif{L}{\x{k}}}
+\dots+\hdi{k-1}(b)\pdif{L}{\x{k}}+\\
+\intgd{a}{b}h^i\br{\pdif{L}{x^i}-\dif{}{t}\pdif{L}{x_{(1)}^i}+\dif{^2}{t^2}\pdif{L}{\x{2}}-\dots+(-1)^k\dif{^k}{t^k}\pdif{L}{\x{k}}}dt,
}
where in the expressions which are not in the integrand all derivatives of $L$ are taken in the point $\br{\g^i\br{b},\dots,\gdi{k}\br{b}}$, and we assume $\dif{}{t}=\x{1}\pdif{}{x^i}+\x{2}\pdif{}{\x{1}}+\dots+\x{2k-1}\pdif{}{\x{2k-2}}$.
Recall now that the well-known formula for the Euler-Lagrange equations for systems with higher derivatives has the form
\eq{
\pdif{L}{x^i}-\dif{}{t}\pdif{L}{x_{(1)}^i}+\dif{^2}{t^2}\pdif{L}{\x{2}}-\dots+(-1)^k\dif{^k}{t^k}\pdif{L}{\x{k}}=0
}
which coincides with the integrand in \re{varfin}. Therefore the integral vanishes and we obtain the final formula
\eqm{
\delta S = h^i(b)\br{\pdif{L}{x_{(1)}^i}-\dif{}{t}\pdif{L}{\x{2}}+\dots+(-1)^{k-1}\dif{^{k-1}}{t^{k-1}}\pdif{L}{\x{k}}}
+\\
+\dot{h}^i(b)\br{\pdif{L}{x_{(2)}^i}-\dif{}{t}\pdif{L}{\x{3}}+\dots+(-1)^{k-2}\dif{^{k-2}}{t^{k-2}}\pdif{L}{\x{k}}}
+\dots+\hdi{k-1}(b)\pdif{L}{\x{k}}.
}

\end{document}